\documentclass[12pt,preprint]{aastex}

\newcommand{\HII}{H\,{\sc ii} }
\newcommand{\spitzer}{\textit{Spitzer} }
\newcommand{\ISO}{\textit{ISO} }
\newcommand{\msx}{\textit{MSX} }
\newcommand{\iras}{\textit{IRAS} }
\newcommand{\coa}{CO~\textit{J}\,=\,1\,--\,0 }
\newcommand{\cob}{$^{13}$CO~\textit{J}\,=\,1\,--\,0 }
\newcommand{\coc}{C$^{18}$O~\textit{J}\,=\,1\,--\,0 }
\newcommand{\hcop}{HCO$^{+}$~\textit{J}\,=\,1\,--\,0 }
\newcommand{\cod}{CO~\textit{J}\,=\,3\,--\,2 }
\newcommand{\coe}{CO~\textit{J}\,=\,6\,--\,5 }

\newcommand{\csf}{CS~\textit{J}\,=\,5\,--\,4 }
\newcommand{\nha}{NH$_{3}$~(\textit{J},\,\textit{K})\,=\,(1,\,1) }
\newcommand{\nhb}{NH$_{3}$~(\textit{J},\,\textit{K})\,=\,(2,\,2) }

\slugcomment{Accepted by \apj} \shorttitle{Multi-wavelength Study of
S87} \shortauthors{Xue, \& Wu}

\begin{document}
\title{A Multi-wavelength Study of the Massive Star-forming Region S87}


\author{Rui Xue\altaffilmark{1,2} and Yuefang Wu\altaffilmark{1}}
\email{rxue@pku.org.cn, yfwu@bac.pku.edu.cn}

\altaffiltext{1}{Department of Astronomy, Peking University, Beijing
100871, China} \altaffiltext{2}{Current address: National
Astronomical Observatories, Chinese Academy of Sciences, Beijing
100012, China}


\begin{abstract}
This article presents a multi-wavelength study towards the massive
star-forming region S87, based on a dataset of
submillimeter/far-/mid-infrared (sub-mm/FIR/MIR) images and
molecular line maps. The sub-mm continuum emission measured with
JCMT/SCUBA reveals three individual clumps, namely, SMM\,1, SMM\,2,
and SMM\,3. The MIR/FIR images obtained by the \textit{Spitzer Space
Telescope} indicate that both SMM\,1 and SMM\,3 harbor point
sources. The \textit{J}\,=\,1\,--\,0 transitions of CO, $^{13}$CO,
C$^{18}$O, and HCO$^+$, measured with the 13.7\,m telescope of the
Purple Mountain Observatory, exhibit asymmetric line profiles. Our
analysis of spectral energy distributions (SEDs) shows that all of
the three sub-mm clumps are massive (110\,---\,210\,$M_{\odot}$),
with average dust temperatures in the range $\sim$ 20\,---40\,K. A
multi-wavelength comparison convinces us that the asymmetric
profiles of molecular lines should result from two clouds at
slightly different velocities, and it further confirms that the
star-forming activity in SMM\,1 is stimulated by a cloud-cloud
collision. The stellar contents and SEDs suggest that SMM\,1 and
SMM\,3 are high-mass and intermediate-mass star-forming sites
respectively. However, SMM\,2 has no counterpart downwards
70\,$\mu$m, which is likely to be a cold high-mass starless core.
These results, as mentioned above, expose multiple phases of star
formation in S87.
\end{abstract}


\section{INTRODUCTION} \label{sect:intro}

Massive stars play an important role in the evolution of the
interstellar medium (ISM) and galaxies; nevertheless their formation
process is still poorly understood from the observational
perspective because of their relatively short evolution periods,
complex ambient circumstances and gregarious nature. Two important
approaches are systematic surveys and multi-wavelength studies
towards individual sources to increase our knowledge about high-mass
molecular cores that may harbor forming massive stars or mark the
sites of future massive star formation.

Previous surveys of high-mass star-forming regions focused on the
sources associated with ultra-compact (UC) \HII regions and their
precursors (PUCHs) \citep{chu2002}. These works identified a number
of high-mass protostellar objects (HMPOs)
\citep{mol1996,mol2002,sri2002,beu2002,wu2006}. However, the
identified objects usually have high luminosities ($L_{\rm IR} >
10^{3}\,L_{\odot}$), indicating that most of them do not represent
the earliest stage of massive star formation. By comparing
millimeter and mid-infrared (MIR) images of fields containing
candidate HMPOs, \citet{sri2005} further identified a sample of
potential high-mass starless cores (HMSCs), which may be the sites
of future massive star formation. However, their MIR identification
based on 8.3\,$\mu$m images from the \textit{Midcourse Space
Experiment} (\textit{MSX}) was not sufficient to validate a genuine
HMSC, because a heating accreting protostar may remain undetected up
to 8\,$\mu$m \citep{bs2007}. The recently released high-resolution
sensitive MIR and far-infrared (FIR) images obtained by the Galactic
survey of the \textit{Spitzer Space Telescope} could be used to
verify these HMSC candidates.

At the same time, some works towards individual sources indicated
that massive cores at early stages might exist in the vicinity of
evolved star-forming sites like UC \HII or \HII regions
\citep{for2004,gar2004,wu2005}. These works suggest that previously
identified evolved sources may harbor objects at various
evolutionary phases, including HMSCs, high-mass cores harboring
accreting protostars, and HMPOs \citep{beu2007}. One scenario is
that the early-stage objects are stimulated by the star-forming
activities in evolved regions. Another hypothesis is that they may
form with their evolved companions during the fragmentation of
parent clouds, but are restrained to give birth to stars in time by
some physical supporting mechanisms. The third possible explanation
is that the detected objects are just diffuse quiescent gas/dust
clumps and will not form stars eventually. Probing the physical
properties and circumstances of these objects may help to address
the questions above.

S87, cataloged as an optical \HII nebula by \citet{sha1959}, is a
complex star-forming region at a distance of $\sim$ 2.3\,kpc
\citep{racine1968,cra1978}. It is associated with a bright FIR
source IRAS~19442+2427 and has been studied by a number of authors.
\citet{chr1986} detected two 22\,GHz water masers in S87.
\citet{bar1989} studied it at radio, infrared, and optical
wavelengths, suggesting the existence of a biconical outflow. A
compact \HII region was detected in centimeter radio continuum, with
an extended emission component \citep{bal1983,bar1989}. Two
near-infrared (NIR) clusters were identified by \citet{che2003},
labeled as S87E and S87W. The submillimeter (sub-mm) continuum
emission of S87 exhibited an asymmetric spatial configuration
\citep{Jen1995,hun2000,mue2002}, which was also confirmed by the
molecular line map of \csf \citep{shi2003}. The previous works in
ammonia (NH$_{3}$) lines \citep{zin1997,stu1984} exposed two
kinematically separate components, which spatially overlap in the
direction of S87E. The recent work of \citet{sai2007} identified
several gas clumps in the \coc map and proposed a hypothesis that
S87E was formed by a cloud-cloud collision. All of the works
mentioned above suggest that complex spatial and kinematic
structures exist in S87, which may harbor objects at different
evolutionary phases. The abundant data currently available from
various wavelengths give us a great opportunity to perform a further
comprehensive investigation towards S87. It may construct a
consistent physical picture for this massive star-forming region and
test the previously proposed hypotheses.

In this article, we present a multi-wavelength study of S87, mainly
based on the online archival data, our observations in molecular
lines, and two published observations \citep{zin1997,mue2002}. We
describe the used dataset and the observational results of S87 in
\S\,\ref{sect:obsdata} and \ref{sect:results}. We concentrate on the
spectral energy distribution (SED) analysis of the identified sub-mm
clumps in \S\,\ref{sect:SED}. In \S\,\ref{sect:discussion}, we
mainly discuss the stellar content, the star-forming activity and
evolutionary stage of each sub-mm clump. The conclusions are
summarized in \S\,\ref{sect:conclusion}.

\section{DATA AND OBSERVATIONS}
\label{sect:obsdata}

\subsection{Continuum Data}

All of the sub-mm/FIR/MIR continuum maps or images of S87 were
obtained from data archives.

The 850 and 450\,$\mu$m sub-mm continuum data were retrieved from
the James Clerk Maxwell Telescope\footnote{The James Clerk Maxwell
Telescope is operated by the Joint Astronomy Centre on behalf of the
Science and Technology Facilities Council of the United Kingdom, the
Netherlands Organisation for Scientific Research, and the National
Research Council of Canada.} (JCMT) Science Archive, measured with
the Submillimetre Common-User Bolometer Array (SCUBA)
\citep{hol1999} installed at JCMT. Two 850 and 450\,$\mu$m maps are
available since S87 was observed twice in 2003. One observation was
carried out in jiggle map mode on 2003 May 24th (JCMT program ID:
M03AN23); the other was performed in Emerson II scan map mode on
August 24th (M03BU45). The beamwidths of JCMT were 7\arcsec.5
(450\,$\mu$m) and 14\arcsec (850\,$\mu$m). All of the retrieved data
have been fully calibrated with the ORAC-DR pipeline \citep{jen2002}
for flat-fielding, extinction correction, sky noise removal,
despiking, and removal of bad pixels, in the units of mJy
beam$^{-1}$.

The \spitzer MIR/FIR data were retrieved from the \spitzer Science
Center\footnote{\url{http://ssc.spitzer.caltech.edu}}, including the
3.6, 4.5, 5.8, and 8.0\,$\mu$m images measured with the Infrared
Array Camera (IRAC) \citep{faz2004} and the 24 and 70\,$\mu$m images
measured with the Multiband Imaging Photometer for \spitzer (MIPS)
\citep{rie2004}. The IRAC and MIPS data are, respectively, from the
Galactic Legacy Infrared Mid-Plane Survey Extraordinaire (GLIMPSE)
\citep{ben2003} and the recently released MIPS Inner Galactic Plane
Survey (MIPSGAL) \citep{car2005}. All of them were calibrated by the
\spitzer Science Center data processing pipelines. In addition, we
also retrieved the MIR images and point source catalog (PSC) of \msx
\citep{ega2003} from the Infrared Processing and Analysis Center
(IPAC)\footnote{\url{http://www.ipac.caltech.edu}} for our study.

\subsection{Spectral Observation}

To investigate the molecular gas of S87, we mapped a region of
4\arcmin~$\times$~4\arcmin~ centered on IRAS~19442+2427 in the
\textit{J}\,=\,1\,--\,0 transitions of CO, $^{13}$CO, C$^{18}$O and
HCO$^+$, with the 13.7\,m millimeter telescope of the Purple
Mountain Observatory (PMO) in 2005 January and 2006 May. A cooled
SIS receiver was employed, and the system temperature $T_{\rm sys}$
at the zenith was $\sim$ 250\,K (SSB). The backend included three
acousto-optical spectrometers, which was able to measure the
\textit{J}\,=\,1\,--\,0 transitions of CO, $^{13}$CO, and C$^{18}$O
simultaneously.  All the observations were performed in position
switch mode. The center reference coordinates are:
R.A.\,(J2000)\,=\,19$^{\rm h}$46$^{\rm m}$19$^{\rm s}$.9,
Dec.\,(J2000)\,=\,+24\degr35\arcmin24\arcsec. The grid spacings of
the CO and HCO$^+$ mapping observations were 60\arcsec and
30\arcsec~respectively. The background positions were checked by
single point observations before mapping. The pointing and tracking
accuracy was better than 10\arcsec. The obtained spectra were
calibrated in the scale of antenna temperature $T_{\rm A}^{*}$
during the observation, corrected for atmospheric and ohmic loss by
the standard chopper wheel method \citep{ku1981}. Table~\ref{tb1}
summarizes the basic information about our observations, including:
the transitions, the center rest frequencies $\nu_{\rm rest}$, the
half-power beam widths (HPBWs), the bandwidths, the equivalent
velocity resolutions ($\Delta V_{\rm res}$), and the typical rms
levels of measured spectra. All of the spectral data were
transformed from the $T_{\rm A}^{*}$ to $T_{\rm mb}$ scale with the
main beam efficiencies before analysis. The uncertainty of
brightness was estimated as 10\%. The
GILDAS\footnote{\url{http://www.iram.fr/IRAMFR/GILDAS}} software
package (CLASS/GREG) was used for the data reduction \citep{gl00}.

\subsection{Other Archival Data}

We acquired the 350\,$\mu$m continuum and \nha line maps of S87
through private communications with K. Young and I. Zinchenko. The
350\,$\mu$m map was measured with the Sub-mm High Angular Resolution
Camera (SHARC) installed at the Caltech Sub-mm Observatory (CSO)
\citep{mue2002}. The \nha line map was obtained with the Effelsberg
100\,m telescope \citep{zin1997}. The technical details are
summarized in the corresponding reference articles.

\section{RESULTS}
\label{sect:results}
\subsection{Sub-mm Maps}
\label{sect:dec}

Fig.~\ref{fig1} displays the 850\,$\mu$m scan map (contours) and the
\spitzer 8.0\,$\mu$m image (inverse grey-scale), in which the NIR
clusters S87E and S87W are revealed as two bright MIR nebulae. The
strongest peak of 850\,$\mu$m is associated with S87E, and two other
peaks exist to the northeast of it. We propose that these three
850\,$\mu$m peaks are associated with three individual sub-mm
clumps. They lie along an axis from southwest to northeast,
hereafter labeled as SMM\,1, SMM\,2, and SMM\,3.

We processed the sub-mm maps using the Richardson-Lucy (RL)
iteration deconvolution algorithm to moderately enhance the spatial
resolutions. As many other deconvolution solutions, this algorithm
did not produce uncertainty information of the results. Therefore,
we have to note that this process is not targeted to get the most
``accurate" deconvolved maps. Our steps of deconvolutions are
similar to those described by \citet{smi2000}. The 850 and
450\,$\mu$m beam patterns of SCUBA were constructed from the Uranus
maps measured in 2003 August. A procedure in
Starlink/KAPPA\footnote{\url{http://www.jach.hawaii.edu/software/starlink/}}
was used to perform the image processing tasks. We avoided the
pixels at the edge of sub-mm maps during the iterations due to their
low signal-to-noise level.

The deconvolutions of the two 850\,$\mu$m maps converged within 100
iterations and produced acceptable enhanced maps without apparently
artificial structures. However, for the 450\,$\mu$m maps, the
procedure failed to converge within 150 iterations. Fig.~\ref{fig2}
displays the deconvolved 850\,$\mu$m maps and undeconvolved
450\,$\mu$m maps. All of them have been converted into the units of
mJy arcsec$^{-2}$. SMM\,3 is not covered in the jiggle maps (see
Fig.~\ref{fig2}c and Fig.~\ref{fig2}d) due to the limitation of the
observational fields of view. SMM\,1 is clearly elongated in the
deconvolved 850\,$\mu$m maps, and there are extended lobes to the
west and south of its peak. SMM\,2 is slightly elongated in the
north-south direction. All of the three sub-mm clumps are revealed
in a common envelop, suggesting that they may be associated although
not necessarily in the same sky plane.

To evaluate the \cod contribution to the 850\,$\mu$m data, we
examined our previous observation of S87 in \cod at the K\"{o}ln
Observatory for Sub-mm Astronomy (KOSMA) 3m telescope. This
observation was carried out for a CO multi-line survey of (UC) \HII
regions and has not been published yet (Xue \& Wu~2008, in
preparation). After converting the integrated intensity of \cod to a
flux density at 850\,$\mu$m, we found that the contribution of \cod
is less than the noise level of the 850\,$\mu$m maps. We also
evaluated the contribution of \coe to the 450\,$\mu$m data by
estimating its integrated intensity from \cod, under the assumption
of local thermodynamic equilibrium (LTE) with an excitation
temperature of 80\,K. We found that its contribution is also small.
Therefore, the effect of line contaminations can be ignored at 450
and 850\,$\mu$m.

\subsection{Mid/Far-Infrared Images}
\label{sect:mir}

\begin{figure}
\centering
\includegraphics[angle=0,scale=.8]{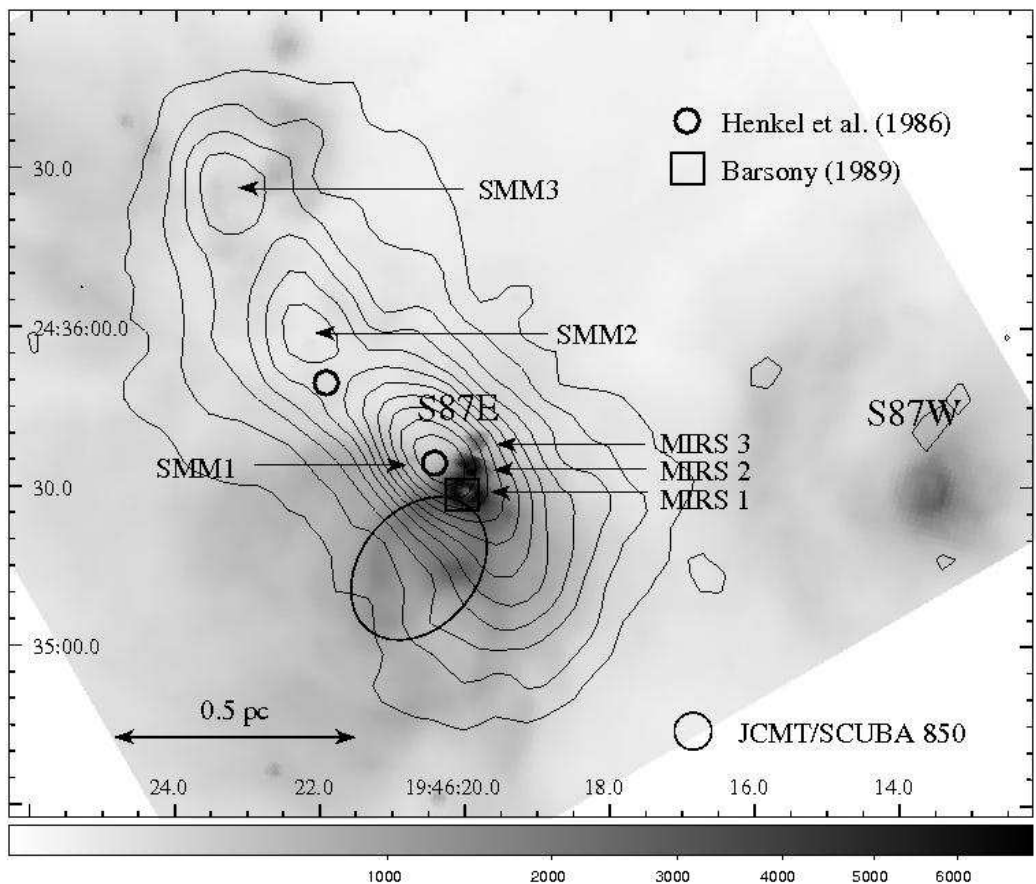} \caption{JCMT/SCUBA 850\,$\mu$m
continuum emission (contours) overlaid on the \textit{Spitzer}/IRAC
8.0\,$\mu$m image. The contour levels increase from 0.6 to 5.0\,Jy
beam$^{-1}$, in a step of 0.4\,Jy beam$^{-1}$ (4$\sigma$). The
inverse grey-scale 8.0\,$\mu$m image is in the unit of MJy
sr$^{-1}$. The coordinate system is J2000. The two small open
circles mark the water masers detected by \citet{chr1986}. The
square denotes the compact \HII region, and the open ellipse
indicates the position and rough size of the associated extended
centimeter emission \citep{bar1989}. The open circle at the bottom
shows the beamwidth of JCMT at 850\,$\mu$m. The sub-mm clumps
SMM\,1, SMM\,2, and SMM\,3, as well as the NIR clusters S87E and
S87W, are labeled. Three 8.0\,$\mu$m point sources, named as
MIRS\,1, MIRS\,2, and MIRS\,3 in \S~\ref{sect:mir}, are marked by
the arrows. \label{fig1}}
\end{figure}

\begin{figure}
      \centering
      \includegraphics[angle=0,scale=.8]{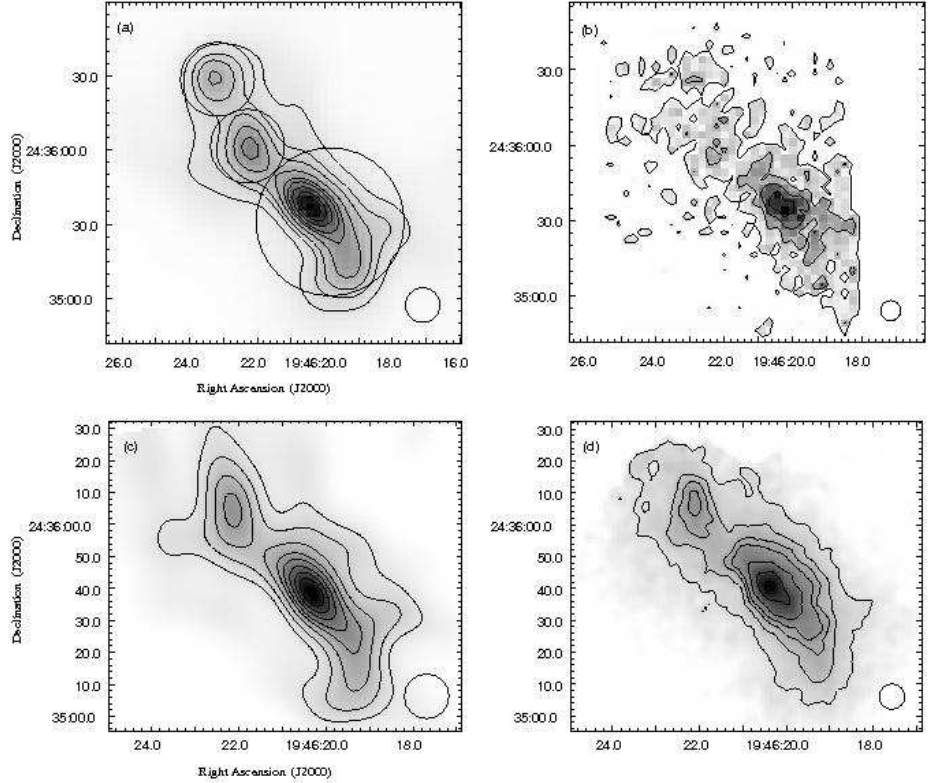}
\caption{Contour plots of the sub-mm continuum maps, overlaid on
their inverse grey-scale images. The beamwidth is indicated in the
bottom-right corner of each panel. (a) The deconvolved 850\,$\mu$m
scan map (M03BU45), after 98 iterations. The contours increase from
1.8 to 29 mJ arcsec$^{-2}$ in a step of 3.1 mJ arcsec$^{-2}$. The
three open circles over the emission peaks denote the photometric
apertures (see \S\,\ref{sect:SED}). (b) The undeconvolved
450\,$\mu$m scan map (M03BU45). The contours are: 40, 120, 220, and
380 mJ arcsec$^{-2}$. (c) The deconvolved 850\,$\mu$m jiggle map
(M03AN23), after 51 iterations. The contours increase from 2 to 26
mJ arcsec$^{-2}$ in a step of 3 mJ arcsec$^{-2}$. (d) The
undeconvolved 450\,$\mu$m jiggle map (M03AN23). The contours are:
60, 120, 200, 280, 360, 440, and 500\,mJ arcsec$^{-2}$.
\label{fig2}}
\end{figure}

\begin{figure}
\includegraphics[angle=-90,scale=.6]{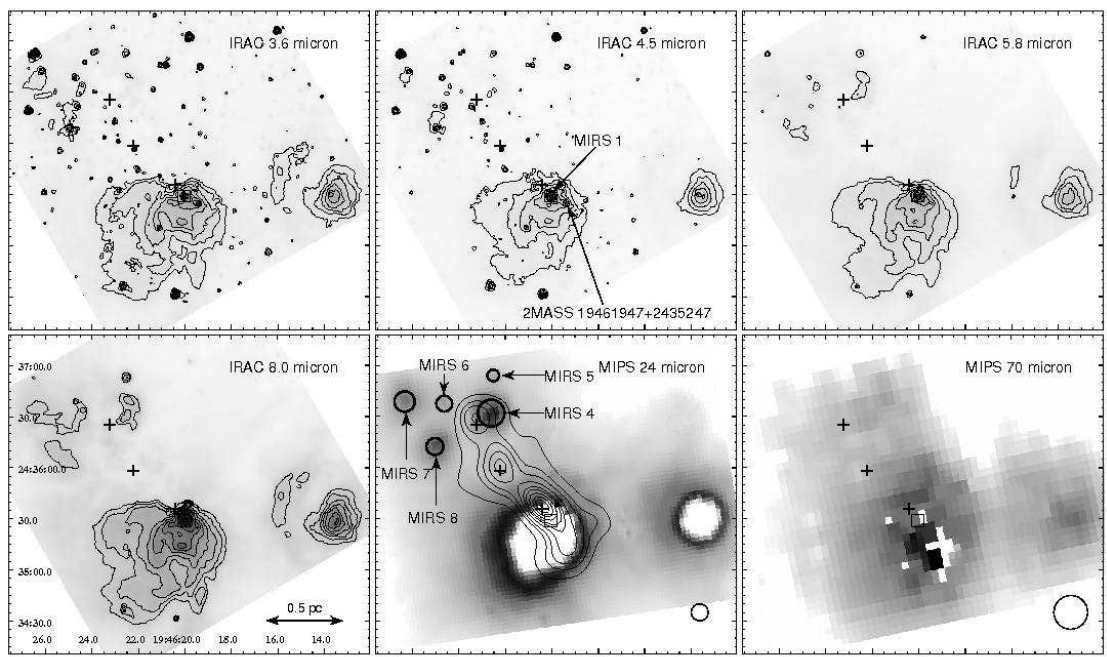}
\caption{The \spitzer MIR/FIR images at six wavelengths. The sizes
and positions of six panels are the same. The absolute coordinates
(J2000) are exhibited in the 8.0\,$\mu$m panel. The white areas in
the emission regions of 24 and 70\,$\mu$m images are due to the
saturation of the detectors. The MIR emission is also plotted as
contours in the 3.6, 4.5, 5.8, and 8.0\,$\mu$m panels. Their contour
levels are: 1\%, 2\%, 3\%, 5\%, 10\%, 16\%, 28\%, 50\%, and 90\% of
the peak intensities. The crosses in each panel denote the
850\,$\mu$m peaks. The square indicates the compact \HII region. The
contour plot of the 24\,$\mu$m panel is the 850\,$\mu$m continuum
emission, adopted from Fig.~\ref{fig2}a. MIRS\,4 to MIRS\,8 are also
marked in the 24\,$\mu$m panel. The small open circles in the
bottom-right corner of 24 and 70\,$\mu$m panels denote the
resolutions of these observations. \label{fig3}}
\end{figure}

A luminous MIR point source is revealed at the position of the
compact \HII region (see the IRAC images of Fig.~\ref{fig1} and
Fig.~\ref{fig3}), which we henceforth label as MIRS\,1\footnote{The
recently released GLIMPSE Spring '05 Catalog names it
SSTGLMC~G60.8838-0.1282}. A weaker MIR point source is found in the
3.6, 4.5, and 5.8\,$\mu$m images, $\sim$ 8\arcsec~to the southwest
of MIRS\,1. It is also detected by the 2MASS NIR All-Sky Survey and
cataloged as 2MASS~19461947+2435247. However, we did not find the
NIR counterpart of MIRS\,1 in 2MASS images, indicating that MIRS\,1
is highly obscured by the surrounding gas/dust envelope at NIR
wavelengths. In the zoomed 8\,$\mu$m image of Fig.~\ref{fig1}, two
other point sources are found to the north of MIRS\,1, which we
label as MIRS\,2 and MIRS\,3. Strong diffuse MIR emission exists to
the southeast of MIRS\,1, coincident with the extended centimeter
emission detected by \citet{bar1989}. Faint diffuse MIR emission is
detected in the northeast of the IRAC images, coincident with
SMM\,3. SMM\,2 has no MIR counterpart in all of the IRAC bands.

S87E and S87W saturate the 24 and 70\,$\mu$m MIPS images. Five other
sources are detected in the 24\,$\mu$m band, which are coincident
with the diffuse MIR emission in the IRAC bands. We label them as
MIRS\,4 to MIRS\,8 (see Fig.~\ref{fig3}). Although the 24 and
70\,$\mu$m images are saturated towards SMM\,1, it is still clear
that the peaks of sub-mm and 24\,$\mu$m emission are separate, which
is also confirmed by a comparison with the \msx E band
(21.3\,$\mu$m) image. SMM\,3 is associated with MIRS\,4. However,
its sub-mm peak and MIRS\,4 are also slightly separate. There are
only weak emission patches in the IRAC 5.8 and 8.0\,$\mu$m bands
towards SMM\,3, indicating that MIRS\,4 is still embedded in its
gas/dust cocoon. No 24 or 70\,$\mu$m emission is detected towards
SMM\,2, suggesting that it may be less evolved.

\subsection{Molecular Lines}
\label{sect:line}

\begin{figure}
      \centering
      \includegraphics[scale=.7]{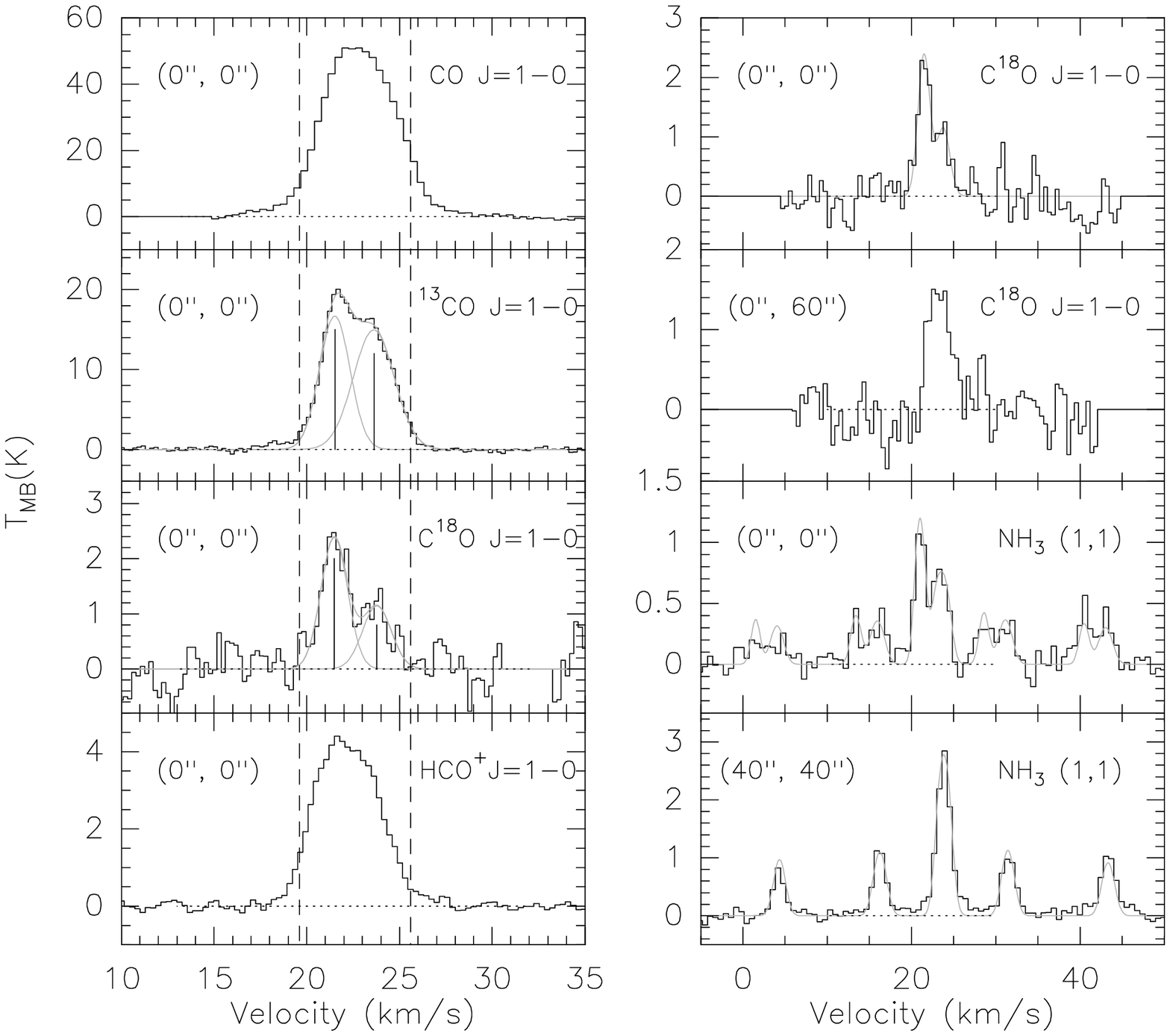}
\caption{{\it Left panel}: The spectra of \coa, \cob, \coc, and
\hcop at the reference coordinates: R.A.\,(J2000)\,=\,19$^{\rm
h}$46$^{\rm m}$19$^{\rm s}$.9,
Dec.\,(J2000)\,=\,+24\degr35\arcmin24\arcsec. The vertical dashed
lines define the border of the \coc emission above 3$\sigma$ level.
{\it Right panel}: The \coc spectra from our observation and the
\nha spectra from \citet{zin1997}. The spatial positions of the
spectra are denoted in the top-left corners of small plots. The thin
lines on the spectra are the fitted results. We performed double
Gaussian fitting for the \cob and \coc spectra. For the \nha
transition, we used the NH$_{3}$ hyperfine structure fitting
procedure of GILDAS/CLASS. \label{fig4}}
\end{figure}

The \textit{J}\,=\,1\,--\,0 transitions of CO, $^{13}$CO, and
HCO$^+$ exhibit asymmetric line profiles (see the left panel of
Fig.~\ref{fig4}), and two components are detected in \coc. Since
\coc is usually optically thin, we can rule out the possibility that
the asymmetric line profile in the other transitions is caused by
the self-absorption in an infall envelope \citep{mye1996,wu2005}.
Previous observations of \citet{stu1984} and \citet{zin1997} also
detected two separate components in \nha and \nhb lines. The \nha
spectra of \citet{zin1997} and our \coc spectra at several positions
are plotted in the right panel of Fig.~\ref{fig4}. These spectra
further confirm that the broad lines of \coa, \cob, and \hcop
consist of two components.

We fitted our spectra at the reference position with Gaussian
profiles. The results are displayed as the thin lines in
Fig.~\ref{fig4}, and the corresponding derived parameters are
summarized in Table~\ref{tb2}, including: the line center
velocities, the fitted line widths, and the brightness temperatures.
We estimated the beam-averaged column densities of C$^{18}$O at the
reference position using the standard LTE method. The excitation
temperature of each component is assumed to be 35\,K, in agreement
with the estimation from \coa (assuming it is optically thick). The
derived C$^{18}$O column densities are $\sim$ 7.8$\times10^{15}$ and
4.0$\times10^{15}$\,cm$^{-2}$ for the components at low and high
velocities.

Fig.~\ref{fig5}a is the \hcop position-velocity diagram along the
northeast-southwest direction, which also exhibits two components.
One is located at the reference position and associated with
SMM\,1; 
the other extends from the reference position to the northeast,
coincident with SMM\,2 and SMM\,3. We propose that these components
arise from two clouds. Hereafter, they are named Cloud I and II,
corresponding with the components at low and high velocity
respectively.

The integrated intensity maps of \hcop and \nha are also exhibited
in Fig.~\ref{fig5}. Two different integrated intervals are adopted,
chosen to separate the emission from Cloud I and II. All of the
presented intensity maps suggest: SMM\,2, SMM\,3, and the northeast
part of SMM\,1 may be associated with Cloud II; the main part of
SMM\,1 is contributed by Cloud I.

\section{SED ANALYSIS}
\label{sect:SED}

\subsection{Observational SEDs}

We extracted the 850 and 450\,$\mu$m flux densities of each clump
using a photometric procedure in the Starlink/GAIA software package.
The measured results, as well as the positions and sizes of the
adopted photometric apertures, are summarized in Table~\ref{tb3}. We
note that the uncertainties in Table~\ref{tb3} are just statistical
errors (rms deviations derived from clean regions), and the
estimation of the overall photometric uncertainties is difficult due
to the limited information from the online data archives. However, a
comparison among different observational modes and the previous
similar observation may provide an evaluation of the accuracy of our
results.

\citet{Jen1995} observed S87 using the receiver UKT14 at JCMT in
1994. They detected two sources, which were coincident with SMM\,1
and SMM\,2 respectively. Our photometric results at 850\,$\mu$m are
in good agreement with theirs (see the last two columns of
Table~\ref{tb3}), but the 450\,$\mu$m results from SCUBA are
systematically larger. Since UKT14 is a single-element bolometer and
its measurements may be affected by the change of sky conditions and
other factors, we believe that the calibration of SCUBA data is more
reliable. The photometric differences of the jiggle and scan maps
are acceptable, less than 20\% at 450\,$\mu$m.

We examined the CSO map and the MIPS images to measure the flux
densities of each clump at 24, 70 and 350\,$\mu$m. Since SMM\,1
saturates the 24 and 70\,$\mu$m images, only lower limits can be
derived at these wavelengths. In addition, we checked the \msx PSC
and found that the photometric apertures of SMM\,1 and SMM\,3 are
coincident with the \msx point sources MSX6C~G060.8828-00.1295 and
MSX6C~G060.9049-00.1275 respectively. Their flux densities are also
adopted to construct the SEDs of SMM\,1 and SMM\,3.

The measured flux densities are summarized in Table~\ref{tb4},
extending from sub-mm to MIR. The average results of the scan and
jiggle maps are adopted for 850 and 450\,$\mu$m. Their differences
are considered as the uncertainties. The 350\,$\mu$m uncertainties
follow the description of \citet{mue2002} and the uncertainties at
24 and 70\,$\mu$m are the statistical errors.

\subsection{Isothermal Dust Model}

A simple isothermal gray-body dust model is used to fit the
observational SEDs. The details follow the method described by
\citet{sch2007}. In the adopted model, the mean weight of
interstellar materials per hydrogen molecule $\mu$ is $\sim$ 2.33.
The dust opacity (mass absorption coefficient) $\kappa_{\lambda}$ is
dependent on the wavelength and can be described using the equation:
\begin{equation}
\kappa_{\lambda} = 
\kappa_{1300} (\frac{1300\,\mu\rm m}{\lambda})^{\beta}, \label{eq4}
\end{equation}
where $\beta$ is the dust opacity index and $\kappa_{1300}$ is the
dust opacity at 1300\,$\mu$m. Assuming a gas-to-dust ratio of 100,
we adopt $\kappa_{1300} = 0.009\,$cm$^2$\,g$^{-1}$, which is derived
from a gas/dust model with thin ice mantles \citep{oh1994}.

We used a non-linear least-squares method (the Levenberg-Marquardt
algorithm coded within IDL) to obtain the best-fit models for the
observational SEDs. The physical properties of each sub-mm clump
were obtained, including: the average dust temperature $T_{\rm d}$,
the dust opacity index $\beta$, and the aperture-average column
density of hydrogen molecules $N_{\rm H_{2}}$. In this fitting test,
only the data upwards 70\,$\mu$m were used. We assumed the
70\,$\mu$m flux density of SMM\,1 to be 4000\,Jy, which was
estimated from the interpolation of the IRAS flux densities
subtracted with the potential contributions from SMM\,2 and SMM\,3.

Fig.~\ref{fig6} displays the best-fit model SEDs for three clumps.
We further calculated their clump masses and bolometric luminosities
$L_{\rm SED}$ (by integrating the model SEDs over the range
1\,$\mu$m$<\lambda<2.0$\,mm). All of these derived results are
summarized in Table~\ref{tb5}. Additionally, we adopted a Monte
Carlo method used by \citet{sch2007} to estimate the errors of the
derived parameters that arise from the observational uncertainties.
The 3$\sigma$ intervals are denoted as the superscripts and
subscripts in Table~\ref{tb5}.

The results in Table~\ref{tb5} show that the sub-mm clumps are all
massive (110\,---\,220\,$M_{\odot}$). SMM\,1 has a higher dust
temperature, and its bolometric luminosity dominates in the whole
region, implying the existence of strong internal heating source(s).
The fitted dust opacity indices of three sub-mm clumps are slightly
different ($\sim$ 1.3\,---\,1.8) and consistent with the typical
values between 1 and 2 \citep{hil2006}. It must be noted that the
derived $N_{\rm H_2}$ is directly affected by the adopted value of
$\kappa_{1300}$. If we reduce $\kappa_{1300}$ by a factor of 2,
$N_{\rm H_2}$ and the derived clump mass $M$, will increase by a
factor of 2. However, the other derived parameters will not be
affected by this change.

\subsection{Two-temperature Dust Model}

In Fig.~\ref{fig6}, the best-fit models of SMM\,1 and SMM\,3 failed
to describe the observational results below 70\,$\mu$m. However, the
model SED of SMM\,2 can explain the absence of its MIR emission. To
better characterize the excess MIR emission of SMM\,1 and SMM\,3, we
performed another SED fitting test using a model with two dust
components at different temperatures. In this fitting test, we
adopted the observational data upwards 14.7\,$\mu$m (excluding
24\,$\mu$m for SMM\,1). To reduce the fitting parameters, we assumed
$\beta$ is 1.5 and 1.3 for each dust component of SMM\,1 and SMM\,3
respectively. The best-fit model SEDs are exhibited in
Fig.~\ref{fig7}, and the derived parameters of the warm and cool
dust components are listed in Table~\ref{tb6}.

The two-temperature model fits the observational data very well
above 12\,$\mu$m, which is consistent with the physical fact that
there are warm dust around the internal heating sources and
relatively cool dust envelopes surrounding the star-forming sites in
SMM\,1 and SMM\,3. Although the \iras 100\,$\mu$m flux density
exceeds the model SED of SMM\,1 (see Fig.~\ref{fig7}), we believe
that the deviation is due to the large beam of \iras. The results in
Fig.~\ref{fig7} and Table~\ref{tb6} show that the warm components
contribute little to the total masses and the flux densities at
sub-mm wavelengths, but are required to explain the excess at MIR
wavelengths.

We tried to modify $\beta$ to fit the emission below 12\,$\mu$m.
However, no satisfying results were found. The emission in the \msx
A and C bands does not follow the predication of gray-body models,
suggesting these models are invalid at these wavelengths. Generally,
two significant spectral features may exist at this MIR wavelength
range. One is the emission of polycyclic aromatic hydrocarbons
(PAHs), which is often detected towards compact \HII regions and
photodissociation regions (PDRs). Previous studies have showed that
the \msx A and C bands often contain PAH emission lines
\citep{gho2002,kra2003,pov2007}. The other is the silicate feature,
which has been predicted in the dust model of \citet{oh1994} and
demonstrated to be important in the recent sophisticated SED models
\citep{rob2006,rob2007}. This feature may be expected as the
absorption at 9.7\,$\mu$m towards some UC \HII regions, produced by
the dust cocoons around center objects \citep{fai1998}.
\citet{pee2002} have identified both of PAH and silicate features
towards S87E in the previous \ISO spectroscopy observation. The
silicate absorption feature may be caused by Cloud II, which partly
overlaps above the compact \HII region. Since our gray-body models
are focused to evaluate the overall properties of dust clumps, which
are mainly constrained by the thermal emission from longer
wavelengths, a detailed SED model explaining the PAH and silicate
features is beyond our purpose.

\section{DISCUSSION}
\label{sect:discussion}

\subsection{Cloud-Cloud Collision}
\label{sect:cc}

All of the FIR/sub-mm images and molecular line maps exhibit the
complex spatial and kinematical structures of S87. The recently
published high-resolution \coc observation \citep{sai2007} revealed
several gas clumps at different velocities, which further confirms
our identification of Cloud I and II. However, are Cloud I and II
really related to each other? \citet{sai2007} proposed that the gas
clumps at higher velocity might be on the near side along the
line-of-sight because the observation of \citet{che2003} detected
many reddened sources in NIR there. They further reasoned that the
clumps at low and high velocities were approaching and the NIR
cluster S87E was possibly formed by a cloud-cloud collision. In the
following, we verify the cloud-cloud collision model by a
multi-wavelength comparison.

Firstly, the 8.0\,$\mu$m emission shows a sharp edge to the
northeast of MIRS\,1 (see Fig.~\ref{fig1}). This feature is probably
caused by the large extinction at 8.0\,$\mu$m because the sub-mm
emission is still strong. The position of the extinction patch is
consistent with that of Cloud II, confirming that Cloud II is on the
near side along the line-of-sight. Therefore, Cloud I and II are
approaching.

\begin{figure}
      \centering
      \includegraphics[angle=0,scale=.6]{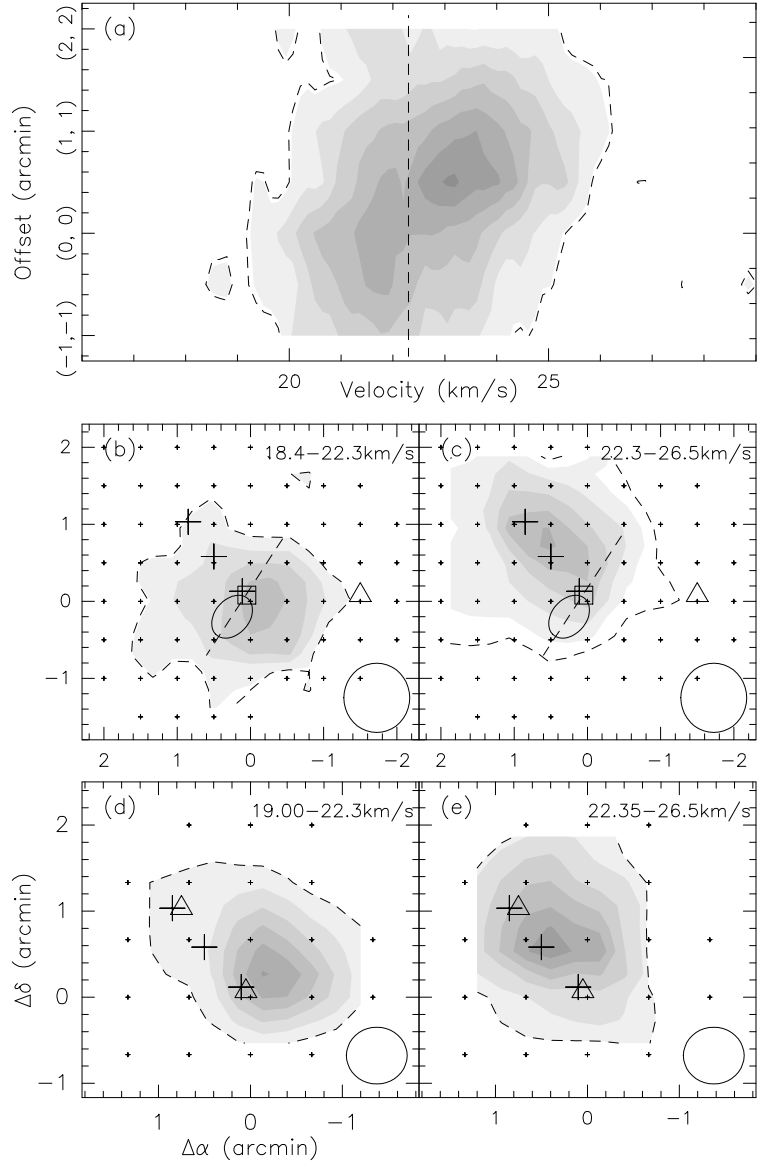}
\caption{In each plot, the dashed contour denotes the 3$\sigma$
level. The center reference coordinates are:
R.A.\,(J2000)\,=\,19$^{\rm h}$46$^{\rm m}$19$^{\rm s}$.9,
Dec.\,(J2000)\,=\,+24\degr35\arcmin24\arcsec. (a): The \hcop
position-velocity diagram along the northeast-southwest direction.
The contour levels increase from 3$\sigma$ to 21$\sigma$ by
3$\sigma$. The vertical dashed line indicates the border of two
integrated intervals (see in \S\,\ref{sect:line}). (b) and (c): The
integrated intensity maps of \hcop. The integrated interval is
showed in the top-right corner of each plot. The contour levels
increase from 40\% to 100\% by 15\% (5$\sigma$) of the peak value.
The square denotes the compact \HII region, and the open ellipse
indicates the position and rough size of the associated extended
centimeter emission \citep{bar1989}. The crosses denote the
850\,$\mu$m peaks, and the beamwidth is indicated in the
bottom-right corner of each plot. The dashed line is the location of
the potential contact plane mentioned in \S\,\ref{sect:cc}. The
position of S87W is marked by triangles. (d) and (e): The integrated
intensity maps of \nha. The contour levels increase from 3$\sigma$
in a step of 3$\sigma$. The spectra data are from \cite{zin1997}.
The positions of MIRS\,1 and MIRS\,4 are marked as triangles.
\label{fig5}}
\end{figure}

\begin{figure}
\includegraphics[angle=-90,scale=.57]{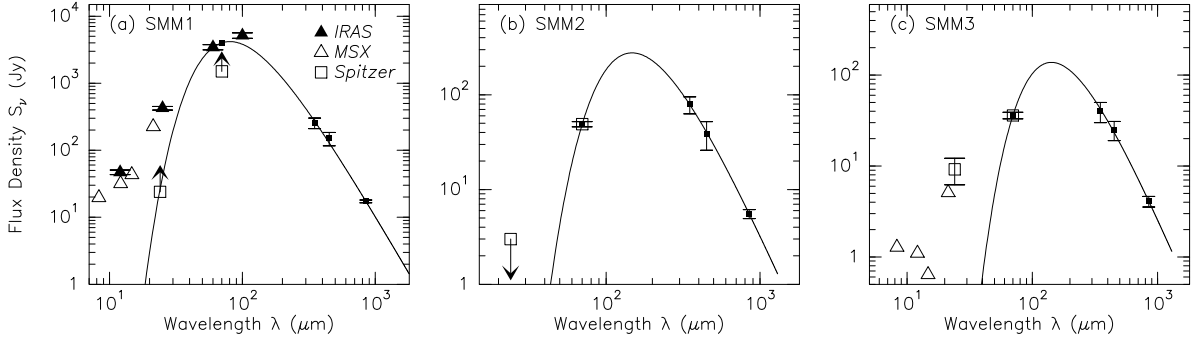}
\caption{From (a) to (c), the best-fit SEDs (solid lines) of SMM\,1,
SMM\,2, and SMM\,3 are exhibited. The filled triangles denote the
flux densities of IRAS~19442+2427. The open triangles denote the
flux densities extracted from \msx PSC. The open squares indicate
the flux densities derived from the \textit{Spitzer}/MIPS images.
The small filled squares on the model SEDs are the data points used
for the SED fitting. The errorbar of each data point is plotted if
available. \label{fig6}}
\end{figure}

\begin{figure}
\centering
\includegraphics[angle=-90,scale=.57]{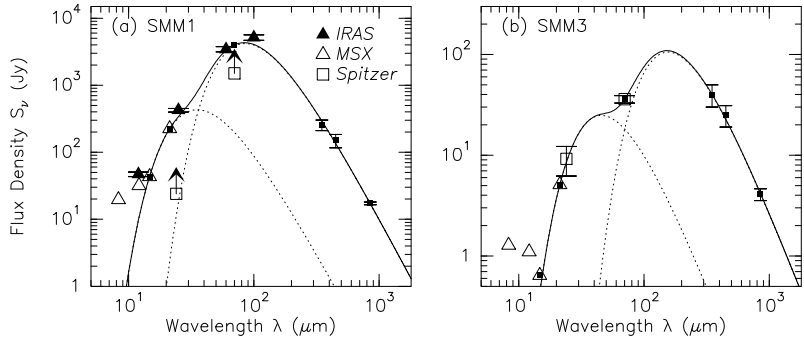}
\caption{The panel (a) and (b) exhibit the best-fit SEDs of SMM\,1
and SMM\,3 (solid lines). The two components of model SEDs are also
displayed (dot lines) respectively. The filled triangles denote the
flux densities of IRAS~19442+2427. The open triangles denote the
flux densities extracted from \msx PSC. The open squares indicate
the flux densities derived from the \textit{Spitzer}/MIPS images.
The small filled squares on the model SEDs are the data points used
for the SED fitting. The errorbar of each data point is plotted if
available. \label{fig7}}
\end{figure}

Next, we can infer from the intensity maps of Fig.~\ref{fig5} that
the peak of SMM\,1 and S87E are in the overlapping region of Cloud I
and II. The MIR point sources in SMM\,1 suggest that there is not
only a formed NIR cluster, but also ongoing star-forming activities.
The strong and continuous star-forming process is likely to be
interpreted by the stimulation of a cloud-cloud collision rather
than the spontaneous evolution of molecular clouds alone.

Furthermore, the spatial configuration of the compact \HII region
and the associated extended centimeter emission
\citep{bar1989,kur1994} also supports the cloud-cloud collision
model. The champagne flow model of \citet{kim2001} combined with
clumpy structures of molecular clouds can explain the extended
centimeter component which stretches to the southeast of the compact
\HII region. If Cloud I and II are in contact, their contact plane
will be along the northwest-southeast direction (see
Fig.~\ref{fig5}b and Fig.~\ref{fig5}c). Consequently, the compact
\HII region will be better confined in the direction perpendicular
to the contact plane and the champagne flow should be easier to
spurt out in the southeast direction. If Cloud I and II are not in
contact, the champagne flow will be more likely to splash in the
direction perpendicular to the border of the parent cloud of the
compact \HII region. The observational result is consistent with the
predication of the first scenario, supporting that the two clouds
are colliding. Assuming that the two clouds have typical sizes $R$
$\sim$ 1\,pc and a velocity separation $\delta v$
$\sim$\,2\,km\,s$^{-1}$, the collision duration is at least
$R/\delta v \sim 5\times10{^5}$\,yrs, comparable with the time scale
forming a compact \HII region.

The cloud-cloud collision is considered as an efficient mechanism to
trigger star formation. It may compress molecular gas and lead to
local gravitational collapse \citep{lor1976,hab1992,mar2001}.
However, its possibility is small in the diffuse molecular clouds
\citep{elm1998}. Additionally, high velocity off-axis collisions
could be destructive rather than lead to gravitational instabilities
\citep{hau1981,gil1984}. Therefore, the fraction of star formation
triggered by cloud-cloud collisions may be small in our Galaxy. All
of the current evidence demonstrates that S87 is a new example of
cloud-cloud collisions, and similar samples are still limited
\citep{lor1976,dic1978,koo1994,val1995,buc1996,sat2000,loo2006}.

\subsection{Molecular Line Emission}

\subsubsection{\hcop}

Our observation shows that the line profile of \hcop is similar to
that of \coa (see the left panel of Fig.~\ref{fig4}). Additionally,
we found that both of \coa and \hcop spectra show slight features of
high-velocity (HV) gas when compared with \coc, suggesting that
HCO$^{+}$ extends in diffuse gas rather than simply concentrates in
the dense parts of gas clumps. Previous observational and
theoretical works have pointed out the abundance enhancement of
HCO$^{+}$ in diffuse or shocked gas \citep{tur1995,gir1999}, which
can explain our finding.

However, the formation mechanism of the HV gas in S87 is unclear. We
propose three different explanations: (i) the HV gas may arise from
stellar outflows; (ii) it may be contributed by the high-pressure
shocked material that is squirted out when the clouds collide; (iii)
or, it is from the non-impacting portions of the colliding clouds
since they do not slow down to a common speed during the cloud-cloud
collision. Although \citet{bar1989} identified HV blue and red wings
in her \coa observation and proposed that the HV gas resulted from a
biconical outflow with a wide opening angle viewed at large
inclination, our identification of two individual clouds apparently
rejects this model.
High-resolution and sensitive observations are required to clarify
the origin of the HV gas.

Because both stellar outflows and cloud-cloud collisions can produce
HV gas and broad non-Gaussian line profiles, it is possible that
some observational results previously interpreted as bipolar
outflows are caused by cloud-cloud collisions. However, since the
possibility of cloud-cloud collisions is not high, similar cases
like S87 should be rare.

\subsubsection{\nha}

A feature of the \nha intensity maps is that the NH$_{3}$ emission
tends to ``evade'' the luminous MIR sources. The \nha peak of Cloud
I is separate from MIRS\,1 and the sub-mm peak of SMM\,1. The \nha
emission is absent to the southeast of MIRS\,1, where the diffuse
MIR emission is strong. The \nha peak of Cloud II is coincident with
SMM\,2, which has no MIR counterpart. In contrast, the observations
of \citet{sai2007} and \cite{shi2003} showed that the \coc and \csf
emission is strong in SMM\,1 and SMM\,3. Since both of SMM\,1 and
SMM\,3 are dense clumps identified from sub-mm continuum, their
relatively weak NH$_{3}$ emission may be explained by the
underabundance of NH$_{3}$. \citet{tur1995n} suggested that NH$_{3}$
could be destroyed by C$^{+}$ that dominates in PDRs. However,
molecules like C$^{18}$O are formed via C$^+$ and not affected by
the photo-destruction process \citep{jan1995}. The diffuse 5.8 and
8.0\,$\mu$m emission near SMM\,1 and SMM\,3 is usually contributed
by PAHs and interpreted as a tracer of PDRs. The existence of MIR
emission there, as well as the strong \coc emission and the weak
NH$_{3}$ emission, are consistent with the prediction of the
chemical process proposed in previous works.

\subsubsection{virial States}

The line widths of molecular spectra are usually used to probe the
kinematics of gas clumps. Since Cloud I and II can be well resolved
in NH$_{3}$ lines, we estimate the virial masses of these two clouds
in this section.

We derived the line widths and brightness temperatures of \nha at
the NH$_{3}$ peaks of Cloud I and II, using the hyperfine structure
fitting procedure of GILDAS/CLASS. The results are exhibited as thin
lines in Fig.~\ref{fig4}. The angular diameters $\theta_{\rm obs}$
of Cloud I and II are calculated using the equation:
\begin{equation}
\theta_{\rm obs} = 2\sqrt{\frac{\Omega}{\pi}},
\end{equation}
in which, $\Omega$ is the measured angular area of each cloud. After
that, we corrected the beam effect and estimated the intrinsic sizes
of Cloud I and II following the equation:
\begin{equation}
R = D\frac{\sqrt{\theta_{\rm obs}^2-\theta_{\rm mb}^2}}{2},
\end{equation}
where $R$ is the radius of the gas cloud in pc, $D$ is the distance
of S87, and $\theta_{\rm mb}$ is the beamwidth of the \nha
observation. Assuming that Cloud I and II are homogeneous spherical
gas clouds with a density distribution $\rho \propto r^{-\alpha}$
($\alpha=1.5$) and neglecting the contributions from magnetic field
and surface pressure, the virial masses can be derived using the
equation \citep{mac1988}:
\begin{equation}
M_{\rm vir} = 126(\frac{5-2\alpha}{3-\alpha}) R \Delta V_{\rm
FWHM}^{2},
\end{equation}
in which, $M_{\rm vir}$ is the virial mass in $M_{\odot}$ and
$\Delta V_{\rm FWHM}$ is the full width at half-maximum intensity
(FWHM) of \nha in km~s$^{-1}$.

All the measured and derived parameters of Cloud I and II are listed
in Table~\ref{tb7}, including: the positions of NH$_{3}$ peaks, the
angular and intrinsic sizes, the line widths at NH$_{3}$ peaks, and
the derived virial masses of two clouds. The total virial mass of
Cloud I and II is $\sim$\,430\,$M_{\odot}$, which is much smaller
than the previous estimation ($\sim$\,1080\,$M_{\odot}$) obtained
with the total line width of two components \citep{zin1997} but
comparable with the mass estimated from the SED fitting
($\sim$\,460\,$M_{\odot}$, from the isothermal dust model). However,
we note that the above comparison of the masses estimated from
different approaches can be affected by the adopted dust opacity and
the assumption of $\alpha$. Although a variation of $\alpha$ is not
likely to cause much change in virial masses, the dust opacity may
change by at least a factor of 2 \citep{oh1994}, which leads to a
large uncertainty in the masses estimated from SEDs.

\subsection{Stellar Contents of SMM\,1 and SMM\,3}
\label{sect:smm}

The position of MIRS\,1 is consistent with that of the compact \HII
region, within the astrometric error (1.5\arcsec), indicating that
MIRS\,1 is the exciting massive (proto)star. We examined the
high-resolution centimeter map of \citet{bar1989} and found that
neither MIRS\,2 nor MIRS\,3 shows compact radio continuum emission.
The possible explanation is that MIRS\,2 and MIRS\,3 are less
evolved compared with MIRS\,1 or they are not massive enough to
ionize their surroundings and to excite compact \HII regions.
\citet{chr1986} detected a strong water maser in SMM\,1, which is
often considered to be associated with HMPOs. The velocity range of
this water maser is 21\,---\,25\,km\,s$^{-1}$, in good agreement
with the systematic velocity of the molecular clouds. All the
evidence mentioned above supports that SMM\,1 is a high-mass
star-forming site that harbors massive forming stars or cluster.

The Lyman continuum radiation from massive stars mainly escapes in
the form of free-free emission. \citet{kur1994} estimated that the
Lyman continuum photon flux required to keep the entire region of
S87E ionized was $3.2\times10^{46}$\,photons\,s$^{-1}$, which
corresponds to that of a B0.5 ZAMS star. The bolometric luminosity
of such stars is $\sim$ $3.0\times10^{4}$\,$L_{\odot}$
\citep{cro2005}, slightly smaller than that of SMM\,1 ($\sim$
$3.7\times10^{4}$$L_{\odot}$, from the two-component model). The
extra luminosity of SMM\,1 may come from the relatively weak MIR
sources near MIRS\,1, which can not be traced by the free-free
emission.

SMM\,3 contains the bright 24\,$\mu$m source MIRS\,4. The ratio of
the luminosities from its cool and warm components is $\sim$ 2.1,
lower that of SMM\,1. Its bolometric luminosity is $\sim$
740\,$L_{\odot}$, also lower than that of SMM\,1, which indicates
that SMM\,3 is more likely to be an intermediate-mass star-forming
site.

\subsection{Physical Properties of SMM\,2}
\label{sect:com}

No MIR point source or diffuse emission below 70\,$\mu$m is detected
towards SMM\,2. Since only a cold dust component can describe its
observational SED, strong internal heating sources are not likely to
exist in SMM\,2.

\citet{chr1986} detected a weak 22\,GHz water maser near SMM\,2,
which usually arises from the dense circumstellar disks around
protostars \citep{par2007} or originates in outflows from the birth
of a massive star \citep{van1998}. Since this water maser is in the
velocity range 8\,---\,15\,km\,s$^{-1}$, significantly different
from that of the molecular clouds, we favor the second explanation
for its origin. We notice that this water maser is on a sub-mm
emission ridge connecting the peaks of SMM\,1 and SMM\,2 rather than
near the peak of SMM\,2, and its position uncertainty is large when
compared with sub-mm observations. Therefore, we doubt that this
weak maser is produced by the intrinsic factors of SMM\,2. For
instance, the potential outflows from massive protostars of SMM\,1
may shock the ambient molecular gas of SMM2 and produce a weak water
maser at the rear side of SMM2. This scenario is consistent with the
lower velocity of the water maser. Therefore, we believe that the
existence of this water maser does not necessarily contradict
SMM\,2's physical properties derived from the SED and MIR image
analyses. Based on the information available, we support that SMM\,2
is probably a HMSC that may form massive stars or intermediate star
clusters eventually.

\section{CONCLUSIONS}
\label{sect:conclusion}

We have carried out a multi-wavelength study of the massive
star-forming region S87. The main results are summarized as follows.

1.~We identified three sub-mm clumps in S87, labeled as SMM\,1,
SMM\,2, and SMM\,3. They are estimated to have masses of 210, 140,
and 110\,$M_{\odot}$, with average dust temperatures of 41, 21, and
24\,K respectively (from the isothermal gray-body model).

2.~We examined molecular line maps from our observations and
compared them with previous results of other authors. We concluded
that the star-forming activities in SMM\,1 are stimulated by a
cloud-cloud collision.

3.~We found that HCO$^{+}$ can trace diffuse gas and NH$_{3}$ may be
destructed by chemical processes in the region harboring MIR sources
or exhibiting strong diffuse MIR emission.

4.~We calculated the virial masses of the two colliding clouds,
which are in good agreement with those estimated from SEDs.

5.~The stellar contents and star-forming activities of sub-mm clumps
are identified. Their SEDs reveal that these clumps are at various
evolutionary stages. SMM\,1 and SMM\,3 are high-mass and
intermediate-mass star-forming regions respectively. SMM2 is massive
and cold, has no MIR counterpart, which is probably a HMSC. All of
these results expose that the star formation in S87 is at multiple
phases.

\acknowledgments

We are grateful to K. Young and I. Zinchenko for their sharing of
the 350\,$\mu$m and \nha maps. We would like to thank the staff at
the Qinghai Station of PMO for their assistance during the
observations and Weilai Gu for her help to obtain the \hcop data. We
acknowledge the anonymous referee for his/her careful reading and
helpful suggestions. Qifeng Yin and Sophia Day are also thanked for
their help on the manuscript. This research was funded by the Grant
10733030 and 10128306 of NSFC. It employed the facilities of the
Canadian Astronomy Data Center operated by the National Research
Council of Canada with the support of the Canadian Space Agency, and
it is also partly based on observations made with the
\textit{Spitzer Space Telescope}, which is operated by the Jet
Propulsion Laboratory, California Institute of Technology under a
contract with NASA.

{\it Facilities:} \facility{JCMT}, \facility{Spitzer},
\facility{MSX}

\clearpage

\begin{deluxetable}{rcccccc}
\tabletypesize{\footnotesize}
\tablecaption{Observation Parameters\label{tb1}} \tabcolsep=3pt
\tablewidth{0pt} \tablecolumns{7} \tablehead{
\colhead{Transition} & \colhead{$\nu_{\rm rest}$} & \colhead{HPBW} &
\colhead{Bandwidth} & \colhead{$\Delta V_{\rm res}$} & \colhead{1$\sigma$ rms$^{\rm a}$ }\\
\colhead{} & \colhead{(GHz)} & \colhead{(\arcsec)} & \colhead{(MHz)}
& \colhead{(km s$^{-1}$)} & \colhead{(K chan.$^{-1}$)}} \startdata
\coa &115.271204 &46 &145 &0.37 &0.18\\
\cob &110.201353 &47 &43  &0.11 &0.14\\
\coc &109.782182 &48 &43  &0.12 &0.14\\
\hcop &~~89.188521  &58 &43  &0.26 &0.11\\
\enddata
\tablenotetext{a}{typical value in the scale of $T_{\rm A}^{*}$}
\end{deluxetable}

\begin{deluxetable}{rccccr}
\tablecaption{Parameters of Molecular Lines\label{tb2}}
\tabletypesize{\footnotesize}
\tabcolsep=3pt \tablewidth{0pt} \tablecolumns{6} \tablehead{
\colhead{~~Transition$^{\rm a}$~~} & \colhead{~~$V_{\rm LSR}$~~} &
\colhead{~~~$\Delta V$~~~} & \colhead{~~~$T_{\rm mb}$~~~} \\
& (km s$^{-1}$)  & (km s$^{-1}$) & (K) & } \startdata
$^{13}$CO~\textit{J}\,=\,1\,--\,0$^{\rm 1}$ & 21.51~(0.08) & 1.93~(0.04) & 16.71~(0.38) \\
C$^{18}$O~\textit{J}\,=\,1\,--\,0$^{\rm 1}$ & 21.48~(0.08) & 1.69~(0.19) & ~2.39~(0.39) \\
$^{13}$CO~\textit{J}\,=\,1\,--\,0$^{\rm 2}$ & 23.61~(0.04) & 2.60~(0.06) & 15.04~(0.38) \\
C$^{18}$O~\textit{J}\,=\,1\,--\,0$^{\rm 2}$ & 23.78~(0.16) & 1.83~(0.42) & ~1.14~(0.39) \\
HCO$^{+}$~\textit{J}\,=\,1\,--\,0$^{\rm b}$ & 22.30~(0.04) & 3.10~(0.16) & ~4.54~(0.11) \\
\enddata
\tablenotetext{*}{The error levels are from Gaussian fitting.}
\tablenotetext{a}{The superscripts indicate the different
components.} \tablenotetext{b}{Since the two components can not be
well resolved, the derived parameters are from single Gaussian
fitting.}
\end{deluxetable}

\begin{deluxetable}{ccccrrrrcccc}
\tablecaption{Photometric Results of the Sub-mm Clumps\label{tb3}}
\tabletypesize{\footnotesize}
\tabcolsep=2.5pt \tablewidth{0pt} \tablecolumns{12} \tablehead{
\colhead{Object} & \colhead{$\alpha$} & \colhead{$\delta$} &
\colhead{$\theta_{\rm ap}$} & \colhead{850
$\mu$m$^{\rm a}$} & \colhead{450 $\mu$m$^{\rm a}$} & \colhead{850 $\mu$m$^{\rm b}$} & \colhead{450 $\mu$m$^{\rm b}$} & \colhead{850 $\mu$m$^{\rm c}$} & \colhead{450 $\mu$m$^{\rm c}$}\\
\colhead{} & \colhead{J2000} & \colhead{J2000} & \colhead{(\arcsec)}
& \colhead{(Jy)} & \colhead{(Jy)} & \colhead{(Jy)} & \colhead{(Jy)}
& \colhead{(Jy)} & \colhead{(Jy)}} \startdata
SMM\,1 &19 46 19.8 &+24 35 32   &60 &17.7$\pm$0.1 &133$\pm$10 &16.8$\pm$0.1 &167$\pm$1 &~~17 &110\\
SMM\,2 &19 46 22.3 &+24 36 01   &30 &~5.8$\pm$0.1 &~32$\pm$6~  &~5.3$\pm$0.1 &~45$\pm$1 &~~6.5 &~18\\
SMM\,3 &19 46 23.2 &+24 36 28   &30 &~4.1$\pm$0.1 &~25$\pm$6~  &~~---~~~~~~~  &~~---~~ &~~---~  &~~---~~\\
\enddata
\tablenotetext{a}{flux densities extracted from the scan map
(M03BU45)} \tablenotetext{b}{flux densities extracted from the
jiggle map (M03AN23)} \tablenotetext{c}{flux densities from
\citet{Jen1995}}
\end{deluxetable}

\begin{deluxetable}{crccccrrrr}
\tablecaption{Flux Densities of the Sub-mm Clumps\label{tb4}}
\tabletypesize{\footnotesize}
\tabcolsep=2.5pt \tablewidth{0pt} \tablecolumns{15} \tablehead{
\colhead{Object} & \colhead{850\,$\mu$m} & \colhead{450\,$\mu$m} &
\colhead{350\,$\mu$m} & \colhead{70\,$\mu$m} & \colhead{24\,$\mu$m}
& \colhead{21.3\,$\mu$m$^{\rm a}$} & \colhead{14.7\,$\mu$m$^{\rm
a}$} & \colhead{12.1\,$\mu$m$^{\rm a}$} & \colhead{8.3\,$\mu$m$^{\rm a}$}\\
\colhead{} & \colhead{(Jy)} & \colhead{(Jy)} & \colhead{(Jy)} &
\colhead{(Jy)} & \colhead{(Jy)} & \colhead{(Jy)} & \colhead{(Jy)} &
\colhead{(Jy)} & \colhead{(Jy)}} \startdata
SMM\,1 &17.3$\pm$0.9 &150$\pm$34 &256$\pm$46 &$>1500^{\rm b}$ &$>24^{\rm b}$  &225$\pm$13 &43.3$\pm$2.3  &31.7$\pm$1.6 &19.6$\pm$0.8\\
SMM\,2 &5.5$\pm$0.6  &39$\pm$13  &79$\pm$14  &49$\pm$3   &$<3^{\rm c}$        &---~~~~        &---~~~~        &---~~~~          &---         \\
SMM\,3 &4.1$\pm$0.6  &25$\pm$6   &40$\pm$8  &36$\pm$3   &~9.1$\pm$3.0             &5.1$\pm$0.4    &0.6$\pm$0.1  &1.1$\pm$0.1  &1.3$\pm$0.1\\
\enddata
\tablenotetext{a}{These values are derived from \msx PSC. We adopted
the flux densities of MSX6C~G060.8828-00.1295 and
MSX6C~G060.9049-00.1275 for SMM\,1 and SMM\,3 respectively. SMM\,2
is not detected in all of the four \msx bands.}
\tablenotetext{b}{Only the lower limits are available due to the
saturation of the \spitzer 24 and 70\,$\mu$m images. We assume the
70\,$\mu$m flux density of SMM\,1 to be 4000$\pm$200\,Jy when
performing the SED fitting.} \tablenotetext{c}{no detection}
\end{deluxetable}

\begin{deluxetable}{cccccr}
\tablecaption{Best-fit Parameters for the SEDs (I)\label{tb5}}
\tabletypesize{\footnotesize}
\tabcolsep=2.5pt \tablewidth{0pt} \tablecolumns{6} \tablehead{
\colhead{Object} & $T_{\rm d}$ & $\beta$ & ${N_{\rm H_{2}}}^{\rm a}$ & $M$ & ${L_{\rm SED}}^{\rm b}$ \\
\colhead{} & (K) & & (10$^{22}$\,cm$^{-2}$) & ($M_\odot$) &
($L_\odot$)} \startdata
SMM\,1~~~ &41$^{+9}_{-5}$  &1.5$^{+0.3}_{-0.4}$ &3.2$^{+0.9}_{-0.5}$ & 210$^{+60}_{-40}$ & 31000\\
SMM\,2~~~ &21$^{+2}_{-1}$  &1.8$^{+0.3}_{-0.3}$ &8.4$^{+0.6}_{-0.5}$ & 140$^{+20}_{-30}$ & 1100\\
SMM\,3~~~ &24$^{+2}_{-2}$  &1.3$^{+0.3}_{-0.2}$ &6.4$^{+0.6}_{-0.5}$ & 110$^{+30}_{-30}$ & 600\\
\enddata
\tablenotetext{a}{aperture-average column density.}
\tablenotetext{b}{The luminosities are calculated from the
integration under the model SED curves over the range
$1\,\mu$m$<\lambda<2.0$\,mm.}
\end{deluxetable}

\begin{deluxetable}{lccccr}
\tablecaption{Best-fit Parameters for the SEDs (II)\label{tb6}}
\tabletypesize{\footnotesize}
\tabcolsep=2.5pt \tablewidth{0pt} \tablecolumns{12} \tablehead{
\colhead{Object~~~} & $T_{\rm d}$ & $\beta$ & ${N_{\rm H_{2}}}^{\rm a}$ & $M$ & ${L_{\rm SED}}^{\rm b}$ \\
\colhead{--- component} &(K) & & (10$^{20}$\,cm$^{-2}$) &
($M_\odot$) & ($L_\odot$)} \startdata
SMM\,1--- cool~~  &40  &1.5 &320 &210  &30000  \\
~~~~~~~--- warm~~ &92  &1.5 &0.7  &0.5  &7200  \\
SMM\,3--- cool~~  &23  &1.3 &690 &112  &500    \\
~~~~~~~--- warm~~ &82  &1.3 &0.4  &0.1  &240    \\
\enddata
\tablenotetext{a}{aperture-average column density.}
\tablenotetext{b}{The luminosities are calculated from the
integration under the model SED curves over the range
$1\,\mu$m$<\lambda<2.0$\,mm.}
\end{deluxetable}

\begin{deluxetable}{lccccccc}
\tablecaption{Parameters of virial Mass Estimatation\label{tb7}}
\tabletypesize{\footnotesize}
\tabcolsep=3pt \tablewidth{0pt} \tablecolumns{9} \tablehead{
\colhead{~~Object~~} & \colhead{~~$V_{\rm LSR}$~~} &
\colhead{$\Delta \alpha^{\rm a}$} & \colhead{$\Delta \delta^{\rm
a}$} & \colhead{~~~$\Delta V$~~~} & \colhead{~~~$\theta_{\rm
obs}$~~~} & \colhead{~~~$R$~~~} & \colhead{~~~$M_{\rm vir}^{\rm
b}$~~~} \\
& (km s$^{-1}$)  & (\arcsec) & (\arcsec) & (km s$^{-1}$) & (\arcsec)
& (pc) & ($M_{\odot}$) } \startdata
Cloud I & 21.01(0.06) & 0 & 0 & 1.22(0.15) &  126 & 0.67 & 170 \\
Cloud II & 23.84(0.02) & 40 & 40 & 1.60(0.05) & 116 & 0.60 & 260 \\
\enddata
\tablenotetext{*}{The error levels are from the spectral fitting.}
\tablenotetext{a}{The positions of intensity peaks are listed as
offsets from the reference coordinates: R.A.\,(J2000)\,=\,19$^{\rm
h}$46$^{\rm m}$19$^{\rm s}$.9,
Dec.\,(J2000)\,=\,+24\degr35\arcmin24\arcsec.} \tablenotetext{b}{
The virial mass estimated by \citet{zin1997} is 1080\,$L_{\odot}$,
which was derived from the total width of two components in the \nha
spectra.}
\end{deluxetable}

\clearpage

\clearpage
\thispagestyle{empty}
\setlength{\voffset}{-15mm}
\clearpage
\setlength{\voffset}{0mm}

\end{document}